\documentclass{aa}
\usepackage{times}

\newcommand{\ltsima} {$\; \buildrel < \over \sim \;$}
\newcommand{\simlt}  {\lower.5ex\hbox{\ltsima}}            
\newcommand{\gtsima} {$\; \buildrel > \over \sim \;$}
\newcommand{\simgt}  {\lower.5ex\hbox{\gtsima}}            
\newcommand{\ferg}{erg cm$^{-2}$ s$^{-1}$ }

\newcommand{\be} {\begin{equation}}

\newcommand{\ee} {\end{equation}}

\newcommand{\bc}{\begin{center}}
\newcommand{\ec}{\end{center}}

\def \hcm {\hbox {\ifmmode $ atoms cm$^{-2}\else atoms cm$^{-2}$\fi}}

\def\pdot {\dot P}
\def\ee {1E~1048.1--5937~}

\begin{document}
 
 
\title
{The Anomalous X--ray Pulsar \ee : Phase resolved
spectroscopy with the XMM-Newton satellite}
 
\author{ A.~Tiengo\inst{1,2},  
E.~G\"{o}hler\inst{3}, R.~Staubert\inst{3},  S.~Mereghetti\inst{1}   }			
 
\institute {
{1) Istituto di Fisica Cosmica ``G.Occhialini'',
via Bassini 15, I-20133 Milano, Italy} 
\\
{2) {\it XMM-Newton} Science Operation Center  ,
  Vilspa ESA, Apartado 50727, 28080 Madrid, Spain} 
\\
{3) Univ. T\"{u}bingen, Inst. fur Astronomie und Astrophysik, Abt. Astronomie,
Waldhauserstr.,  D-72076 T\"{u}bingen,    Germany}  
}
 
\offprints{S.Mereghetti, sandro@ifctr.mi.cnr.it}
 
\date{Received / Accepted}
 
\authorrunning{A. Tiengo et al. }
\titlerunning{ \ee}

\abstract{We report on an observation of the Anomalous X--ray Pulsar \ee performed
with the {\it XMM-Newton}   satellite.
The phase averaged spectrum of \ee is well  described by the				
sum of a power law with photon index $\sim$2.9 and a blackbody with temperature
$\sim$0.6 keV, without evidence for significant absorption or emission lines.
The above spectral parameters do not vary during the phases corresponding to
the broad pulse, while the off pulse emission shows a different spectrum
characterized by a soft excess at energies below $\sim$1.5 keV.					
 The {\it XMM-Newton} observation and a re-analysis of  archival {\it BeppoSAX }		
data, show that the spectral parameters and flux of \ee did not change significantly		
during observations spanning the last four years.						
All the data are consistent with a 2-10 keV luminosity varying					
in the range $\sim$(5--7)$\times$10$^{33}$ erg s$^{-1}$  (for a					
distance of 3 kpc).
  }
\maketitle
 
\section{Introduction}

The pulsar \ee is a persistent source of  X-rays with  flux
of $\sim$10$^{-11}$
erg cm$^{-2}$ s$^{-1}$  and a significant
modulation at a period of $\sim$6.45 s (Seward, Charles \& Smale 1986,  Mereghetti 1995,
Corbet \& Mihara 1997, Oosterbroek et al. 1998).
It belongs to the small group of Anomalous X-ray Pulsars
(AXP's, Mereghetti \& Stella 1995, van Paradijs et al. 1995).
The nature of these pulsars, that show properties clearly different
from those of the $\sim$90
X-ray pulsars powered by accretion from high mass companion stars,
is currently unclear (see Mereghetti 2001, for a review).
The lack of evidence for bright optical counterparts led to models for AXP's
involving isolated neutron stars,
but their typical values of  P and $\pdot$ yield rotational energy
losses inadequate to power the observed luminosity.
Although low mass companions cannot be completely ruled out (Mereghetti, Israel \&
Stella  1998), it has been proposed that the AXP's consist of isolated
neutron stars accreting from residual disks (van Paradijs et al. 1995,
Ghosh, Angelini \& White 1997,
Chatterjee  et al. 2000).
Another class of models is   based on isolated, strongly
magnetized (B$>$10$^{14}$ G) neutron stars, or "Magnetars"
(Thompson \& Duncan 1993, 1996;  Heyl \& Hernquist 1997),
powered by the decay of their magnetic energy.

Here we report on a new observation of \ee performed with the
{\it XMM-Newton}   satellite. The new data  
provide an accurate measurement of the   flux, 
phase resolved spectroscopy and 
a more precise  localization of this pulsar.
We have also reanalyzed the {\it BeppoSAX} data originally reported by
Oosterbroek et al. (1998) in order to allow a more accurate comparison
with the new observation.

\begin{table*}[htbp]
\begin{center}
  \caption{Summary of the   Spectral Results$^{(a)}$  }
  
    \begin{tabular}[c]{ccccccc}
\hline
Obs.   & Pow Law      &  Absorption          & kT      & $\chi^2$ / d.o.f. & 2-10 keV Flux    & 2-10 keV   Flux       \\ 
       & photon index &  10$^{22}$ cm$^{-2}$ & (keV)    &                &  absorbed$^{(d)}$   &   unabsorbed (\ferg ) \\
\hline
PN                    &  2.7$\pm$0.3  &  0.9$\pm$0.1  &  0.61$\pm$0.04 & 132.1/129 &  3.8 $\pm$0.4 10$^{-12}$     &   4.3 10$^{-12}$   \\
MOS1                  &  3.1$\pm$0.4  &  1.2$\pm$0.2   &  0.63$\pm$0.07 & 94.7/100 &  4.2 $\pm$0.5 10$^{-12}$     &   5.1 10$^{-12}$   \\
MOS2                  &  3.3$\pm$0.6  &  1.1$\pm$0.2   &  0.71$\pm$0.07 & 100.2/96 &  4.0 $\pm$0.5 10$^{-12}$     &   4.8 10$^{-12}$   \\
EPIC$^{(b)}$          &  2.9$\pm$0.2 &  1.04$\pm$0.08   &  0.63$\pm$0.03 & 340.5/332 &  4.1 $\pm$0.4 10$^{-12}$     &   4.9 10$^{-12}$   \\
\hline
{\it BeppoSAX}$^{(c)}$&  3.3$\pm$0.4 &  1.2$\pm$0.3   &  0.62$\pm$0.04  & 185.8/180 & 5.9$\pm$0.5 10$^{-12}$     &   7.1 10$^{-12}$   \\
\hline

\end{tabular}
\end{center}
\begin{small}
$^{(a)}$ Errors are at the  90\% c.l. for a single interesting parameter\\
$^{(b)}$ EPIC = PN+MOS1+MOS2\\
$^{(c)}$ Reanalysis of the observation reported by Oosterbroek et al. (1988)\\
$^{(d)}$ The errors  on the flux values are dominated by the systematic
uncertainties   (statistical errors are of the order a few percent)
\end{small}
\end{table*}

\begin{figure*}[tb] 
\mbox{} 
\vspace{10.5cm} 
\includegraphics{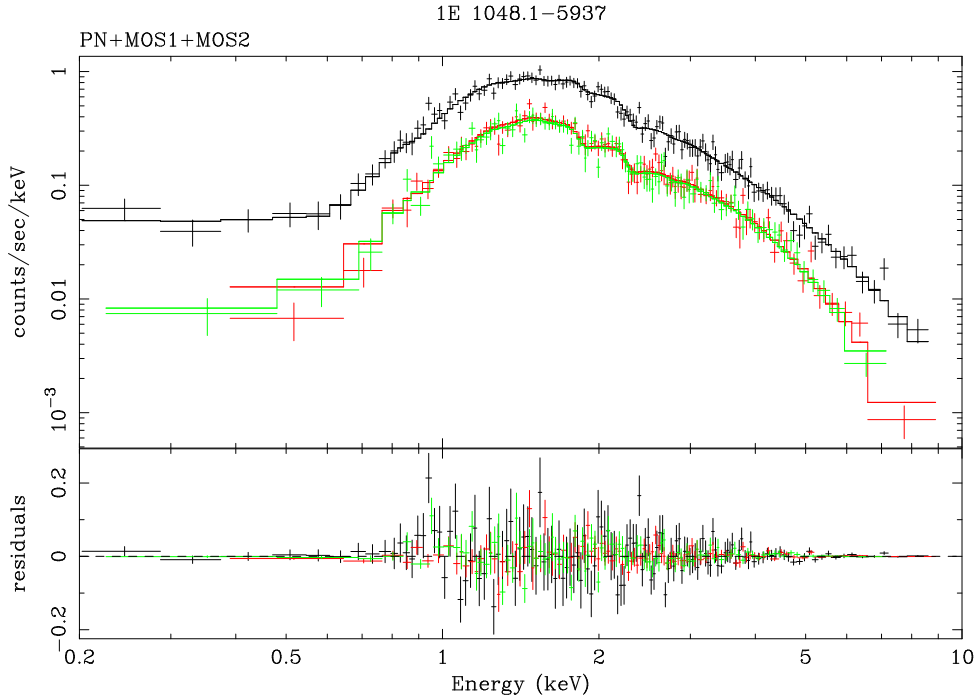} 
 \caption[]{ EPIC spectrum of \ee fitted with the sum of a power law and a blackbody
(see the EPIC parameters in Table 1). The upper spectrum is  from the PN camera. }			
 \label{f1} 
\end{figure*}

\begin{figure*}[tb] 
\mbox{} 
\vspace{10.5cm} 
\includegraphics{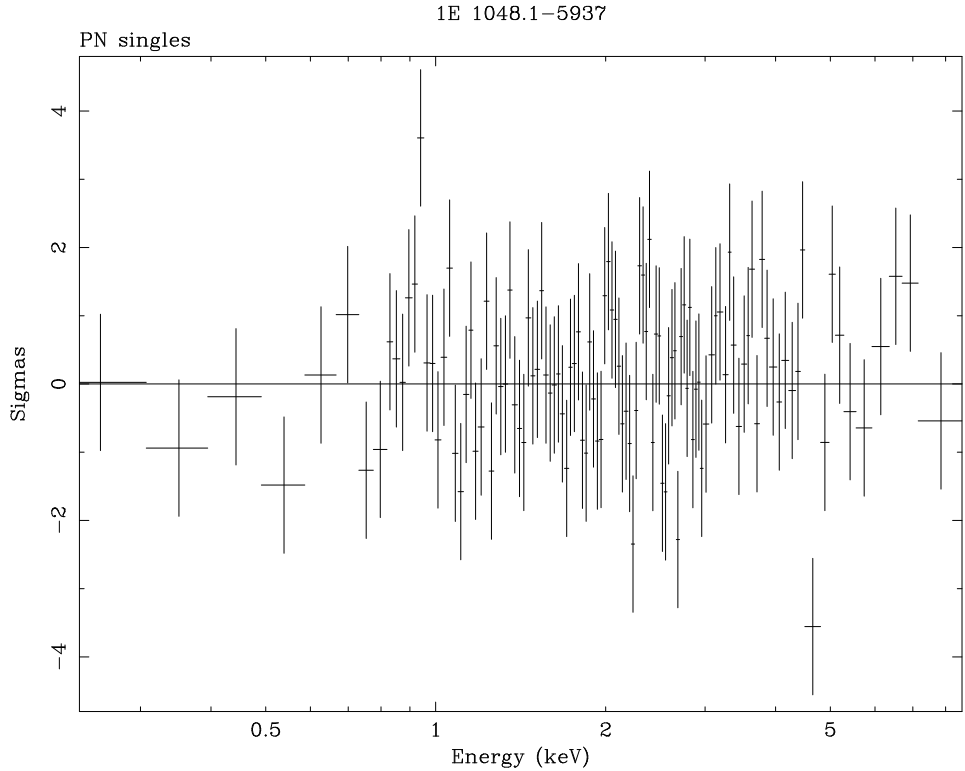} 
 \caption[]{ Residuals of the PN single events spectrum   }
 \label{f2} 
\end{figure*}

\begin{figure*}[tb] 
\mbox{} 
\vspace{10.5cm} 
\includegraphics{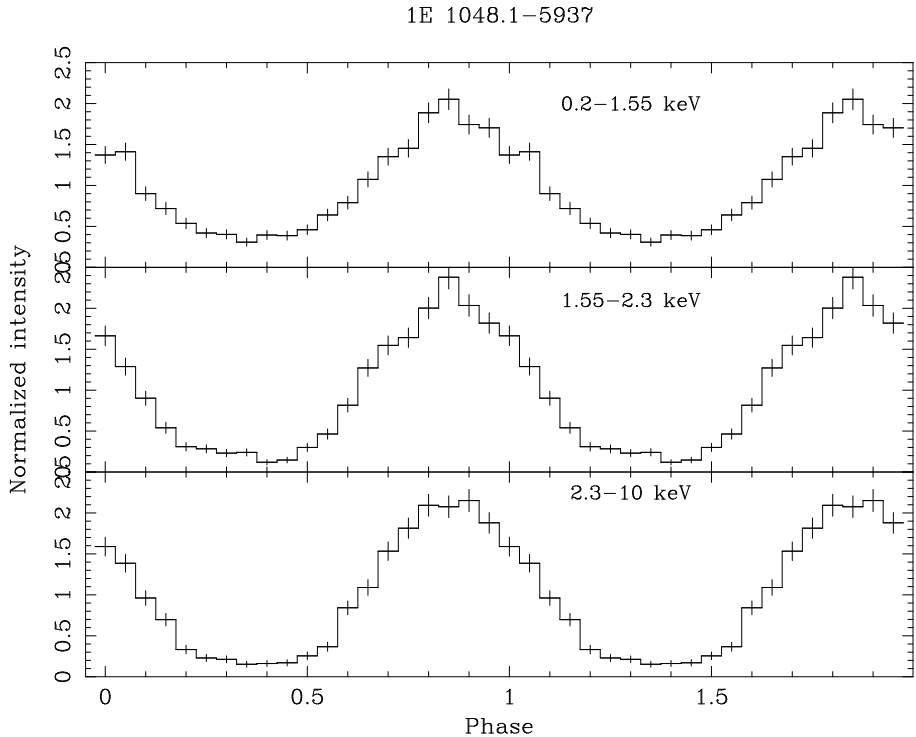} 
 \caption[]{EPIC PN light curves of \ee folded at the best period for 
  different energy ranges}
 \label{f3} 
\end{figure*}

\section{  Observations and Data Analysis  }

An observation of \ee was carried out with the {\it XMM-Newton} satellite
(Jansen et al.  2001) on 2000 December 28. 
Here we concentrate on the  data obtained with the European Photon Imaging Camera 
(EPIC, Turner et al.  2001;
Str\"{u}der et al.  2001).
The short duration of this
observation 
($\sim$8 ks),
prevented a significant detection of \ee in the Reflection
Grating  Spectrometer (RGS). 

EPIC consists of two MOS CCD detectors and
a PN CCD instrument.  In total, the 3 EPICs provide $> 2500$ cm$^{2}$ of
collecting area at 1.5 keV.  The mirror system offers an on-axis FWHM angular
resolution of 4-5$''$ and a field of view (FOV) of 30$'$ diameter.

During this observation the medium filter was used.  The MOS cameras were
operated in Small Window mode to limit source photon pile-up effects and to
improve the time resolution. 
 Due to its shorter frame integration time (73 ms),
the PN could operate in the standard Full Frame mode.
The net exposure time was   7135 s in 
each MOS camera and  4744 s in the PN.

The EPIC data reduction was performed using  version v5.0 of the {\it
XMM-Newton} Science Analysis System.  An additional data cleaning was
required for MOS2 data in order to remove some residual low energy electronic
noise.

\subsection{Spectral Analysis}

The target was clearly detected in the three EPIC cameras at coordinates
R.A.=10$^h$ 50$^m$ 06.3$^s$  Dec. = --59$^{\circ}$ 53$'$ 17$''$ (J2000).
Based on the current performance of the {\it XMM-Newton} satellite attitude
reconstruction we conservatively estimate an uncertainty of $\sim$4$''$ on this position 
that is consistent with previous, less accurate  measurements  (Mereghetti,
Caraveo \& Bignami 1992).

The spectral analysis  with the PN camera was based
on counts extracted from  a  100$''$$\times$100$''$ box
centered at the source position. This box was inside a
single CCD chip.
For the background   extraction we tried several source
free regions close to the target, without finding  significant
differences in the results.  
All the spectra were extracted in the  nominal 0.2-10 keV range and
rebinned so that at least 20 counts were included in each energy bin and the
channels did not oversample the PN energy resolution.   
 
We first analyzed spectra based on both single and double events.
This gives a good statistics without significantly degrading 
the spectral performances.

Single component models, i.e.   power law, thermal bremsstrahlung and
blackbody, gave unacceptable results. A very good fit was obtained
with the ''canonical'' AXP model based on the sum of a 
blackbody and a power law. The best fit
parameters were a blackbody temperature kT$\sim$0.6 keV and a power law 
photon index $\alpha_{ph}\sim$2.7.

The two MOS spectra were based on all the counts of the small CCD   windows
(100$''\times$100$''$) used in this observation. The background
could only be extracted from the peripheral CCD chips, excluding
small regions around sources and correcting the intensity for the
vignetting effect that is significant in this region more than
5$'$  offset from the telescope axis. 
We only considered the 0.2-10 keV energy range and applied the same 
rebinning as in the PN case. 
For both MOS spectra a combination of blackbody and
power law gave the best fit result, with spectral parameters consistent
with those obtained with the PN camera.

We therefore proceeded to a joint spectral analysis of the three data
sets, allowing only the relative normalization to vary.
The results are reported in Table~1 for the blackbody plus
power law model and the corresponding spectrum is shown in Fig.~1.

The   observed flux in the   2-10 keV energy range 
is $\sim$4.1$\times$10$^{-12}$ \ferg 
in the two MOS cameras.
The value obtained with  the PN is 
about 10\% smaller. Part of the discrepancy can be explained 
by the fact that in our
analysis the PN ``Out of Time'' events are excluded. 
They are estimated to be the
6.2\% of the total flux in Full Frame mode. 
The remaining discrepancy is within the uncertainties in the relative
calibration of the instruments.

The best fit spectral parameters reported in Table 1
are consistent with those derived from previous ASCA
observations (Paul et al., 2000), but they show some differences
 with respect to  those observed with {\it BeppoSAX} by					
Oosterbroek et al. (1998), who found a harder powerlaw 						
($\alpha _{ph}$$\sim$2.5) and a smaller absorption 						
($N_{H}$$\sim$0.45$\times$10$^{22}$ cm$^{-2}$).  In order to check if this discrepancy		
is really due to a spectral variation we present a reanalysis of the 
{\it BeppoSAX} data in Section 2.4.

\subsection{Search for spectral features}

To fully exploit the energy resolution provided by the EPIC CCD detectors,
we performed a spectral analysis by selecting only single events. 
Spectra based on single events, i.e. those for which the   collected 
charge is not split between
adjacent CCD pixels, provide the best energy resolution, at the expenses of
a slightly reduced efficiency.

To  extract and rebin source and background spectra from the PN camera
we followed the same procedure described in the previous section. The residuals
of the best fit spectrum (power law plus blackbody) are shown in Fig.~2.
Possible features are seen at energies of $\sim$0.95 keV and in the 
region between $\sim$4 and 5 keV. 

Although the highest point in the low energy feature is at  $\sim$3.5 $\sigma$
from the best fit model, the fact that its width is smaller than the
instrumental resolution at this energy (FWHM $\sim$60-70 eV), suggests
that it is most likely an   artifact. 
 
The absorption at high energy has a width consistent with the instrumental resolution.
It can be fit with a gaussian line 
centered at E=4.71$\pm$0.04 keV and with $\sigma$=45 eV.

None of these two possible   features is confirmed by  the MOS  spectra
(an excess is seen in the residuals at $\sim$0.9 keV only in the MOS2 at  $\sim$2 $\sigma$).

\begin{figure*}[tb] 
\mbox{} 
\vspace{10.5cm} 
\includegraphics{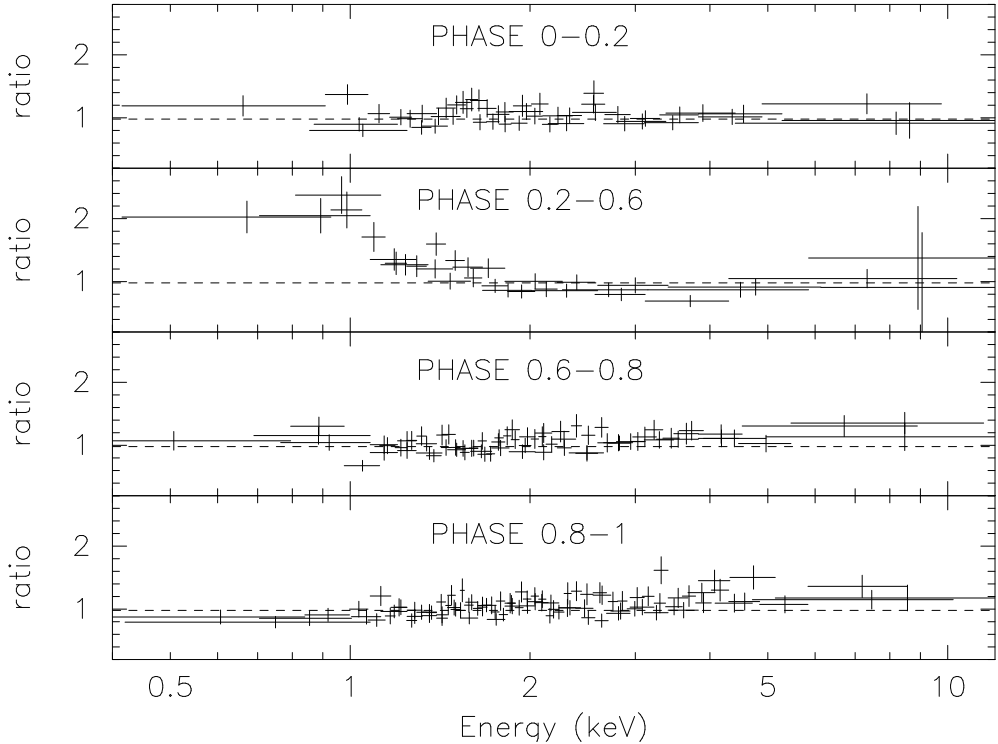} 
 \caption[]{Results of the phase resolved spectroscopy. The figures show the
 ratios between the source spectra in the indicated phase intervals and
 the average spectrum.  
   }
 \label{f4} 
\end{figure*}

\begin{figure*}[tb] 
\mbox{} 
\vspace{10.5cm} 
\includegraphics{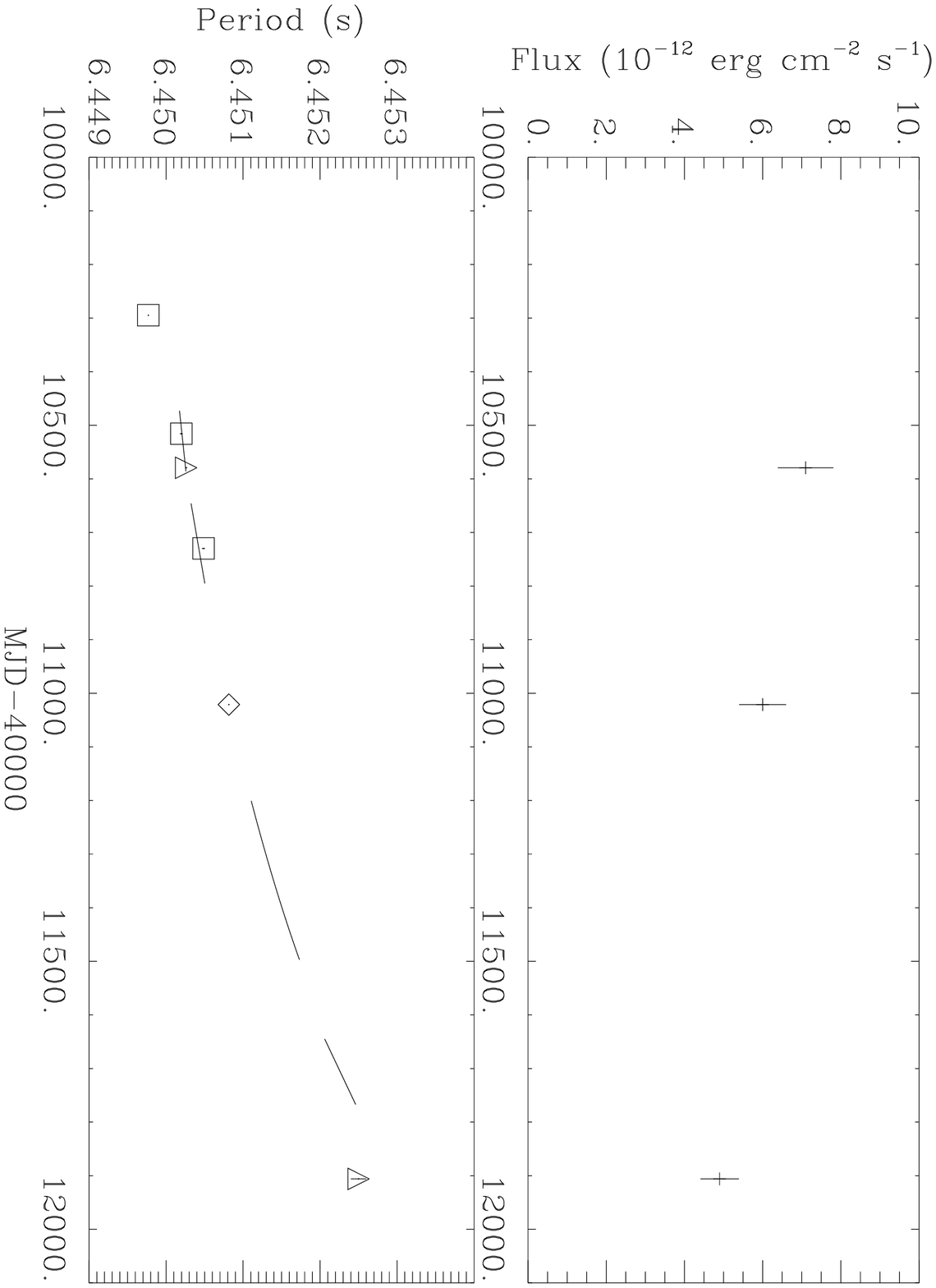} 
 \caption[]{Flux and spin period evolution  of \ee during the years 1996-2000.
The top panel reports the 2-10 keV flux corrected for absorption
as measured with  {\it BeppoSAX}, ASCA and {\it XMM-Newton}.
The different symbols in the bottom panel are as follows. 
Triangles: {\it BeppoSAX} and {\it XMM-Newton} (this work);
Squares: {\it RossiXTE}  (Mereghetti et al. 1998, Baykal et al. 2000);
Diamonds: ASCA (Paul et al. 2000);
Solid lines:  {\it RossiXTE}  (Kaspi et al. 2001).
   }
 \label{f3} 
\end{figure*}

\subsection{Timing analysis and phase resolved spectroscopy}

The timing analysis was done by correcting the time of arrival of
the counts to the Solar System barycenter and then folding them
to several trial periods around the expected value.
The best periods obtained for the PN and MOS data were
respectively  P=6.4526$\pm$0.0001 and 
P=6.4525$\pm$0.0001.  
   
The PN folded light curve is shown 
in Fig.~3 for different energy ranges. 
The pulse profile is characterized by a broad, nearly sinusoidal profile
at all energies, as already noted in previous observations
with other satellites.
The level of modulation was derived by fitting a constant plus a  
sinusoid to the normalized pulse profiles, yielding a pulsed fraction of $\sim$72\%
below 1.5 keV and  $\sim$94\% at higher energies.

To search for possible spectral variations as a function of the spin
phase, we divided the data in four phase intervals
corresponding to the off-pulse (0.2-0.6), rising (0.6-0.8),
top (0.8-1), and declining (0-0.2) parts of the light curve. 
The spectra of the four phase intervals
were fitted with the power law plus blackbody model, 
allowing only the overall normalization to vary and keeping all the other
parameters fixed at the best fit values of Table~1.
The resulting residuals,  see Fig.~4, clearly illustrate that the
average spectral parameters provide a satisfactory fit to the  three
phase intervals covering the rise, peak and decay of the pulse.
On the other hand, the spectrum during the phase interval 0.2--0.6
(corresponding to the minimum in the folded light curve) presents
a significant excess below $\sim$1.5 keV. 

Despite the limited counting statistics, we attempted to model the spectral shape 
in the pulse minimum. A single power law   provides
an acceptable fit with parameters $\alpha_{ph}$$\sim$3.0  and 
N$_H$$\sim$9$\times$10$^{21}$ cm$^{-2}$.
Keeping the absorption fixed at the phase averaged value of 10$^{22}$ cm$^{-2}$,
we obtain $\alpha_{ph}$=3.2$\pm$0.1.

\subsection{Reanalysis of the {\it BeppoSAX}   data}

We reanalyzed the  {\it BeppoSAX} data of the  May 1997 observation
already reported by Oosterbroek et al. (1998).
The main difference with respect to the work of these
authors is our estimate of the background spectrum from a   region
of the field close to the target.
Since \ee lies at low galactic latitude, where significant diffuse X--ray emission
from the Carina nebula is present,  
the use of standard background files from {\it BeppoSAX} observations of
extragalactic empty fields underestimates the background
level (particularly at energies below $\sim$3 keV).

Our extraction regions for the MECS (2-10 keV) were a circle of radius 4$'$ for
the source and a circular corona from 280$''$ to 560$''$ for the background.
The corresponding LECS (0.5-10  keV) values were 8$'$, 520$''$ and 960$''$.
The best fit results to the joint LECS+MECS spectra for the power law plus
blackbody model are given in Table 1.  Both the absorption and the power law
photon index are different from those reported by Oosterbroek et al.  (1998) and
consistent with the EPIC values.  Thus we conclude that there is no evidence for		
a variation in the source spectrum between the two observations. 				
The unabsorbed flux corresponding to the best fit {\it BeppoSAX} parameters is 
7.1$\times$10$^{-12}$ \ferg (2-10 keV), 
 about 45\% higher than that observed with {\it XMM-Newton}.			


\section{Discussion}

The   {\it
XMM-Newton}  observation  of \ee provides the first clear evidence for a spectral 
variation as a function of the pulse phase in this source.
The spectrum shows a ''soft excess'' (with respect to the phase averaged spectrum)
during  the pulse minimum, while no significant
variations occur along the pulse rise and decay. 
This behavior is different from that of other AXP's which show
energy dependent pulse shapes like   1RXS~J170849--4009 (Israel
et al. 2001, Gavriil \& Kaspi 2001) and 4U 0142+61 (Israel et al. 1999). 

Although the spectra of all AXP's are well described by the sum of a 
power law and a blackbody, there are no compelling reasons to 
attribute this spectral shape to the presence of two physically distinct 
emitting components (some suggestions in this 
sense were made, e.g., by Ghosh, Angelini \&  White 1997).
In fact, as noted by \"{Ozel} et al. (2001),
the contribution of the blackbody component to the total
flux is a strong function of  the energy, while the pulsed
fraction at different energies does not change significantly.
In the case of two physically distinct spectral components
this would require an {\it ad hoc} coupling  in the pulse profiles of the
blackbody and power law components.
 
The fact that   the pulsed emission from \ee can be fit by the same
spectral shape at all the phases  also suggests that it originates from  a single physical component,
with a spectrum more complex than a pure blackbody. 
Theoretical AXP's spectra have been  recently computed by several authors
in the context of models of thermal emission from   strongly
magnetized neutron stars. 
These models predict modifications to pure blackbody-like spectra,
in the form of hard power-law tails 
(\"{Ozel} 2001, Perna et al. 2001) and/or narrow spectral features
(Zane et al. 2001).   
Further observations with better statistics are 
clearly required to   confirm the
possible features in the spectrum of \ee reported in section 2.2. 

The period value found in the {\it XMM-Newton} observation
is consistent with the overall spin-down trend at 
an average $\pdot$$\sim$1.6$\times$10$^{-11}$ s s$^{-1}$
that \ee has maintained in the last few years.  
Large variations in the  average spin-down rate were detected in the past.
In particular, between June 1992  and March 1994 (Mereghetti 1995, Corbet \& Mihara 1997),    
$\pdot$$\sim$3.3$\times$10$^{-11}$ s s$^{-1}$ was about a factor 2 higher
than that measured earlier and after 1996.  
More recent observations have clearly shown that the  level of  rotational 
irregularities in \ee is higher than   in other sources of this class 
(Paul et al. 2000, Baykal et al.2000, 2001; Kaspi et al.1999, 2001;  Gavriil \& Kaspi 2001).

In principle, the study of the spin-down evolution in AXP's could
discriminate between accretion-based and magnetar models.
If AXP's are powered by accretion (either  from a companion star or from a residual disk)  
one would expect significant $\pdot$ fluctuations superimposed on the long
term spin-down  trend, possibly correlated with luminosity variations, due to
irregularities in the accretion flow. On the other hand, 
the   period evolution of a magnetar should be much more
regular, with the possible exception of "glitches" as observed in
young radio pulsars.

It is unclear whether the spin-down variations seen in \ee are related
to changes in its X--ray flux, particularly for observations carried out
before 1995.
Oosterbroek et al. (1998) compared the flux of \ee measured with
{\it BeppoSAX} to all the values obtained in the previous
years with the Einstein, EXOSAT, ROSAT and ASCA satellites.
Although the flux values span almost an order of magnitude, the systematic
uncertainties are difficult to evaluate, since some of these
measurements referred to different energy ranges.
Furthermore, the data from non-imaging detectors
might be contaminated by the presence of  the strongly variable   source
Eta Carinae within the field of view.

By means of a systematic program aimed at phase-coherent timing of \ee with the  {\it RossiXTE }
satellite,
Kaspi et al. (1999) showed the
presence of significant $\pdot$ changes between 1996 and 2000.
Their ephemerids are shown by the solid curves in Fig.~5. 
Unfortunately, due to difficulties in the background
subtraction,  the  {\it RossiXTE } non-imaging observations can   
measure only  the intensity of the pulsed flux, which was found
consistent with a constant value.
Our {\it XMM-Newton} and {\it BeppoSAX} results, compared with the ASCA ones (Paul et al. 2000), 
 show that all the  measurements 
obtained since 1994 with imaging detectors are consistent with 
an unabsorbed 2-10 keV flux within the range $\sim$(5-7)$\times$10$^{-12}$ \ferg. 			

Our period measurement is inconsistent with an extrapolation of the phase-coherent
solution reported by Kaspi et al. (2001). This is a further demonstration
of the high level of timing noise present in \ee.

\section{Conclusions}

The large collecting area of the {\it XMM-Newton} telescope has provided 
the first significant evidence for spectral variations as a function of the
spin   phase in \ee. Contrary to other AXP's, the spectral shape changes
only in the off-pulse emission, while no significant variations are seen during
the broad pulse. 

No convincing evidence was found for narrow spectral features. Longer {\it XMM-Newton}
observations (also exploiting the RGS instrument) and improved calibrations 
of the instruments   are required to verify (or not) the structures
marginally detected   in the EPIC PN camera.
 
The {\it XMM-Newton} observation, as well as our re-analysis of the archival {\it BeppoSAX } 
data, show that the spectral parameters and flux of \ee did not change significantly
during observations spanning the last four years. 
The fact that all the observations of good quality obtained with imaging detectors and covering
energies up to $\sim$10 keV are consistent with a 
 2-10 keV luminosity varying in the range $\sim$(5-7)$\times$10$^{33}$ erg s$^{-1}$				
(for d=3 kpc),													
casts some doubts on previous reports of variability up to a factor 10.
 It is unclear whether the large timing noise of \ee can be ascribed to variations in			
the torque exerted on the neutron star by such relatively small fluctuations in the mass accretion rate.	

Finally, we have provided a reduction of a factor $\sim$15 in the area of 
the error box of \ee. This  will undoubtedly help in the search for an optical counterpart of this 
intriguing X--ray source.

\section{Acknowledgments }
 
E.G\"{o}. and R.St. acknowledge the support trough a grant by DLR		
(Verbundforschung).

\end{document}